\documentclass{article}
\usepackage{spconf,amsmath,graphicx}
\usepackage{lipsum}
\usepackage{float}
\usepackage{booktabs}
\usepackage{array}

\def\x{{\mathbf x}}

\newcolumntype{Y}{>{\centering\arraybackslash}X}
\newcolumntype{L}[1]{>{\raggedright\let\newline\\\arraybackslash\hspace{0pt}}m{#1}}
\newcolumntype{C}[1]{>{\centering\let\newline\\\arraybackslash\hspace{0pt}}m{#1}}
\newcolumntype{R}[1]{>{\raggedleft\let\newline\\\arraybackslash\hspace{0pt}}m{#1}}

\title{Emotional Voice Conversion using multitask learning with Text-to-speech}
\name{Tae-Ho Kim$^\star$, Sungjae Cho$^\dagger$, Shinkook Choi$^\dagger$, Sejik Park$^\star$ and Soo-Young Lee$^\star$ 
    \thanks{This work was supported by Institute of Information \& Communications Technology Planning \& Evaluation(IITP) grant funded by the Korea government(MSIT) [2016-0-00562(R0124-16-0002), Emotional Intelligence Technology to Infer Human Emotion and Carry on Dialogue Accordingly], and Ministry of Culture, Sports and Tourism(MCST) and Korea Creative Content Agency(KOCCA) in the Culture Technology(CT) Research \& Development Program 2019.}
}
\address{
    $^\star$KI for Artificial Intelligence, KAIST, Daejeon, Republic of Korea\\
    $^\dagger$Information and Electronics Research Institute, KAIST, Daejeon, Republic of Korea\\
    \texttt{\small \{ktho22,sungjaecho,sk.c,sejik.park,sylee\}@kaist.ac.kr}
}
\begin{document}
%
\maketitle
\begin{abstract}
 Voice conversion (VC) is a task to transform a person's voice to different style while conserving linguistic contents. Previous state-of-the-art on VC is based on sequence-to-sequence (seq2seq) model, which could mislead linguistic information. There was an attempt to overcome it by using textual supervision, it requires explicit alignment which loses the benefit of using seq2seq model. In this paper, a voice converter using multitask learning with text-to-speech (TTS) is presented. The embedding space of seq2seq-based TTS has abundant information on the text. The role of the decoder of TTS is to convert embedding space to speech, which is same to VC. In the proposed model, the whole network is trained to minimize loss of VC and TTS. VC is expected to capture more linguistic information and to preserve training stability by multitask learning. Experiments of VC were performed on a male Korean emotional text-speech dataset, and it is shown that multitask learning is helpful to keep linguistic contents in VC.

\end{abstract}
\section{Introduction}
\label{sec:intro}


Voice can be regarded as the composition of what to deliver --- \textit{linguistic contents} --- and how to deliver --- \textit{style}. Voice conversion (VC) is a task to change speech styles while keeping the linguistic contents \cite{vcgmm1}. It is a challenging task since linguistic contents might be lost or style information is not changed during conversion. 

Previously, VC had been conducted using frame-based approach. Given the source and target speeches, alignment between two speeches is obtained, then acoustic features of source speech are converted to target speech. Various methods were applied to model acoustic features such as Gaussian mixture model (GMM) \cite{vcgmm1, vcgmm2}, deep neural network (DNN) \cite{vcdnn}, recurrent neural network (RNN) \cite{vcrnn}. Recently, successes of sequence-to-sequence (seq2seq) model with attention \cite{bahdanau2014neural} is also applied to VC \cite{vcseq2seq}. Problems such as mispronunciation, training instability have been observed while training seq2seq VC model \cite{vcvip2018}. In \cite{vcvip2018}, textual supervision is added to each time step of decoder output to improve the quality of VC. However, this approach has limitations since it requires explicit alignment by human or dynamic time wrapping (DTW).

To overcome the drawbacks of previous works, we employ an approach to use text-to-speech (TTS). TTS is a task to transform textual or phonetic information to speech waveform, and abundant seq2seq-based studies have actively conducted \cite{taco, taco2, transformertts}. TTS is a highly related task to VC. Only the input domains of VC and TTS are different, the role of the decoder is very same to convert phonetic information to acoustic features. The embedding space of TTS is highly correlated to phonetic information, and  VC is expected to learn embedding space close to that of TTS by using multitask learning. In this paper, TTS is utilized to give phonetic information to VC to improve the performance.

Furthermore, we extend this work to emotional voice conversion. Given a style reference speech, style encoder extracts emotion information only and removes linguistic contents. Since style encoder is designed to extract emotion regardless of linguistic contents, it can handle multiple input style domain. Also, extracted emotion is injected to the decoder, then it can generate various emotions. Thus the proposed model can handle many-to-many emotional voice conversion.

Contributions of our proposed model are as follows:

\begin{itemize}
    \vspace{-5pt}
    \item Multitask learning with TTS could improve the performance of VC.
    \vspace{-5pt}
    \item Many-to-many emotional voice conversion was firstly conducted by a seq2seq model.
    \vspace{-5pt}
    \item A style reference speech could determine target domain of voice conversion.
\end{itemize}

\begin{figure*}[t]
    \centerline{
    \includegraphics[width=15cm]{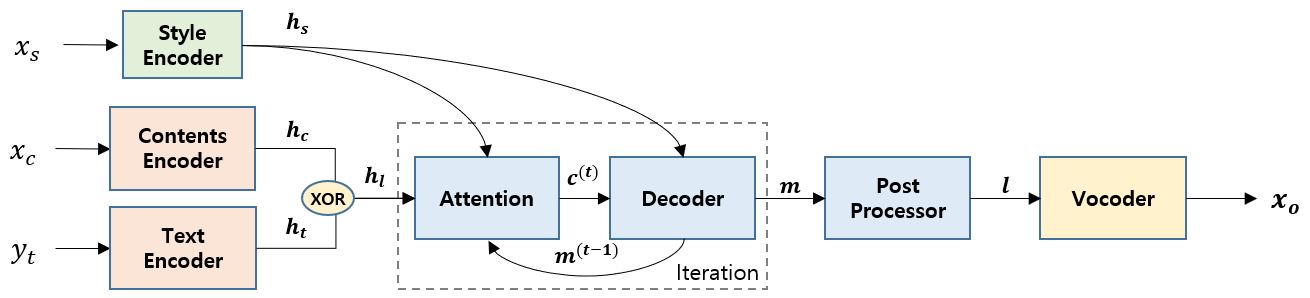}
    }
    \caption{Network architecture of the proposed model.}
    \label{fig:model}
\end{figure*}

The remaining paper is organized as follows: In Section \ref{sec:relatedworks}, related work of our study will be introduced. In Section \ref{sec:methods}, details of proposed methods will be explained. Experimental setup and its results will be shown in Section \ref{sec:results}, and the paper will be concluded in Section \ref{sec:conclusion}.

\section{Related Work}
\label{sec:relatedworks}

Traditionally, frame-based models have been used to solve VC. Given source and target speeches, their alignment is found by DTW. Then conversion between acoustic features of aligned frames are modeled. Recently, seq2seq models have been proposed in VC \cite{vcseq2seq, vcvip2018}. In these models, the model jointly learns alignment and frame conversion by attention mechanism without explicit temporal alignment. However, the performance of VC is not sufficient since converted voice might lose linguistic information. For example, the same words repeatedly generated, some words dropped, or wrongly pronounced. To prevent these phonomena, textual supervision is added to VC, but it requires explicit alignment \cite{vcvip2018}. Unlike previous work, we used TTS for guiding linguistic information to VC. This approach does not need any explicit alignment.

Meanwhile, emotional voice conversion mainly has done with frame-based conversion \cite{LuoCTA17, LuoCTA19} or rule-based approach \cite{xue2018voice}. These have limitations since DTW does not ensure the exact alignment and rule-based approach has a limitation to model voice conversion. It could be improved by using model which does not rely on explicit alignment.

Many-to-many VC refers as to the number of source domain and target domain is multiple. In \cite{keskin2019many}, Cycle-GAN converted an out-of-dataset speaker to a target speaker, and vice versa. The i-vector-based VC system is proposed to generate the linguistic features of speakers that are not in the training set \cite{liu2018voice}. Compared with other many-to-many VC methods \cite{keskin2019many,liu2018voice}, our proposed model can transfer the emotional knowledge of speaker and enables conversion among different emotions.

TTS is one of the most active research area in speech domain. Various kinds of Seq2seq models have been proposed \cite{taco, taco2, transformertts}, and expressive synthesis is also studied \cite{ttsexpressive, lee2017emotional}. Among them, the most relevant work is \cite{lee2017emotional}. Style vector extracted from style encoder which takes a one-hot vector of emotion as an input is injected to the network to generate emotional speech. In our work, emotion labels are not utilized during training, and any emotion labels are not explicitly taken as our network's inputs.

\section{Proposed Model}
\label{sec:methods}

The proposed model can perform VC and TTS in a single model. The network plays as VC when its input is $(\textbf{x}_c,~ \textbf{x}_s)$, or TTS when the input pair is $(\textbf{y}_t, \textbf{x}_s)$ where $\textbf{x}_c,~ \textbf{y}_t$, and $\textbf{x}_s$ are log Mel spectrogram of speech carrying linguistic contents, one-hot represented text, and log Mel spectrogram of style reference speech, respectively. Both $\textbf{x}_c$ and $\textbf{y}_t$ are mapped to the same space $\textbf{h}_l$, then it is decoded into Mel spectrogram $\textbf{m}$. For each decoding step, style vector $\textbf{h}_s$ extracted from $\textbf{x}_s$ is concatenated to attention RNN and decoder RNN. Linear spectrogram $\textbf{l}$ is obtained by the post processor. Detailed network architecture is depicted in Fig. \ref{fig:model} and below equations.

\begin{align}
    \textbf{h}_t &= \text{TextEncoder}(\textbf{y}_t) \nonumber \\
    \textbf{h}_c &= \text{ContentsEncoder}(\textbf{x}_c) \nonumber\\
    \textbf{h}_s &= \text{StyleEncoder}(\textbf{x}_s) \nonumber\\
    \textbf{h}_l &= \text{XOR}(\textbf{h}_c, \textbf{h}_t) \nonumber\\
    \textbf{h}^{(t)}_{att}, \textbf{c}^{(t)}, \textbf{o}^{(t)} &= \text{Attention}(\textbf{m}^{(t-1)}, \textbf{h}_s, \textbf{h}_l, \textbf{h}_{att}^{(t-1)}, \textbf{c}^{(t-1)})\nonumber\\
    \textbf{h}^{(t)}_{dec}, \textbf{m}^{(t)} &= \text{Decoder}(\textbf{c}^{(t)}, \textbf{o}^{(t)}, \textbf{h}_s, \textbf{h}^{(t-1)}_{dec}) \nonumber\\
    \textbf{l} &= \text{PostProcessor}(\textbf{m})\nonumber
\end{align}

where $\textbf{h}_t,~ \textbf{h}_c$, and $\textbf{h}_s$ are the embedding of text encoder, contents encoder, and style encoder, respectively. $\textbf{h}_{att}^{(t)}$ and $ \textbf{h}_{dec}^{(t)}$ are hidden representation of attention RNN and decoder RNN at time step $t$. $\textbf{m},~\textbf{l},~ \textbf{c}^{(t)}$ and $\textbf{o}^{(t)}$ are log Mel spectrogram, log linear spectrogram of target, context vector achieved by attention mechanism, and output of attention RNN at time step $t$. XOR denotes exclusive OR operator.

For the TTS part including text encoder, decoder, attention and post processor, overall architecture is based on Tacotron \cite{taco}, and some modifications that context vector $\textbf{c}^{(t)}$ is utilized for every iteration in attention RNN, and residual connection is added to Convolution Bank + Highway + bi-GRU (CBHG) connection as described in \cite{lee2017emotional}.

Text encoder is composed of character embedding layer, prenet, and CBHG where prenet is composed of two FC-ReLU-Dropout layers. A stack of LSTM is used for contents encoder, and there was no reduction of temporal resolution since temporal reduction may lose local temporal information. It means that the length of $\textbf{h}_c$ is the same as that of $\textbf{x}_c$. For style encoder, $\textbf{x}_s$ is mapped to $\textbf{h}_s$ which has fixed dimension by taking the embedding of the last step of LSTM followed by a fully connected layer to reduce dimension. The attention module is composed of attention RNN followed by attention mechanism. Attention RNN takes input of $\textbf{h}_s,~\textbf{c}^{(t-1)}$ and $\textbf{m}^{(t-1)}$. Then its output $\textbf{o}^{(t)}$ and $\textbf{h}_l$ is used for attention mechanism to generate context vector $\textbf{c}^{(t)}$.

Training loss is $|\textbf{m} - \textbf{m}_{gt}| + |\textbf{l} - \textbf{l}_{gt}|$ where $\textbf{m}_{gt}$ and $\textbf{l}_{gt}$ are ground truth log Mel spectrogram and log linear spectrogram.

\section{Experimental Results}
\label{sec:results}
In this section, description on our dataset and implementation details will be delivered. Experiments are designed to verify enhanced speech quality by multitask learning. Also, the analysis on style encoder and the performance of many-to-many emotional VC will be displayed.

\subsection{Dataset}
We have constructed a male Korean Emotional Text-to-speech (mKETTS) dataset. One 30-year-old male pronounced 3,000 sentences in seven different emotions (neutral, happiness, sadness, anger, fear, surprise, and disgust), so the whole number of utterances is 21,000. The text of all sentences is same across the emotions. All recordings have conducted in a silent studio without
noise, and recorded in 44.1 kHz sampling rate. The whole duration after trimming silence is 29.2 hours.

\begin{table}[t]
\begin{center} 
\caption{WER comparison}
\vskip 0.12in
\begin{tabular}{ccccc}
\toprule
                & VC    & VCTTS-V & VCTTS-T & TTS    \\
\midrule
WER    & 54.54 & 38.50    & 31.98     & 30.39  \\
\bottomrule
\end{tabular}
\vspace{-0.3cm}
\label{table:ling}
\end{center}
\end{table}

\subsection{Implementation details}

For preprocessing, the silence of the first and end of each waveform is trimmed using voice activity detection (VAD) algorithm\footnote{https://github.com/wiseman/py-webrtcvad} and resampled to 16 kHz. Then log Mel spectrogram is extracted with window size 50ms, shift 12.5 ms, nfft 2048, 80 Mel bins and Hanning window. The magnitude was normalized to [0, 1]. Since text encoder uses character-based representation, Korean character decomposed into onset, nucleus, and coda, then converted into one-hot representation. 

For detailed parameter settings are followed. We used 256 character embedding, 32 dimension for $\textbf{h}_s$. The initial learning rate was 1e-3 for Adam optimizer. Gradient clipping was used with 1, and $\textbf{m}$ was generated with reduction factor 5. The batch size was 32, and Bahdanau attention \cite{bahdanau2014neural} was used. For contents encoder, we used two layers of bidirectional LSTM and for style encoder, and contents encoder is composed of two layers of unidirectional LSTM. The output of the last time step is only used for contents encoder. The parameters of contents encoder and style encoder is not shared. In the training stage, teacher forcing is used to prevent accumulating loss of predicted output. When creating mini-batches, each sample is zero-padded to the longest length of the samples. Then losses on the zero-padded regions are also backpropagated for inference stability. For every iteration, the task of the network is randomly decided to either VC or TTS. For each sample,  the source and target emotion are differently selected. 

\begin{table}[t]
\begin{center} 
\caption{MOS and ABX preference score on clarity}
\vskip 0.12in
\begin{tabular}{ccc}
\toprule
                & VC    & VCTTS-V \\
\midrule
MOS             & 4.34 $\pm$ 0.27 & 4.61 $\pm$ 0.28    \\
ABX             & 0.25 $\pm$ 0.14 & 0.59 $\pm$ 0.11    \\

\bottomrule
\end{tabular}
\label{table:mosabx}
\end{center}
\vspace{-0.3cm}
\end{table}

\subsection{Linguistic consistency}

For verifying linguistic consistency of the proposed model, three different models were trained and evaluated. \textbf{VCTTS} is the model that combines TTS and VC together as explained in Section \ref{sec:methods}. \textbf{VC} refers to the voice converter that does not have any TTS path. \textbf{TTS} refers to the TTS model that does not contain VC paths. \textbf{VCTTS-V} is the model for inference of voice conversion using \textbf{VCTTS}, and \textbf{VCTTS-T} is the model when TTS path of \textbf{VCTTS} is activated.

Word error rate (WER) was computed to measure how our proposed model improves the linguistic consistency of the converted speech.
Practically, morphemes were used instead of words since morphemes are considered as recognition units of Korean speech \cite{KwonP03, LeeC04, BangKK18}.
Google Cloud Speech-to-Text API transcribed the converted speech, and transcripts were divided into a sequence of morphemes by the Komoran morphological analyzer in KoNLPy \cite{ParkC14_konlpy}. 
Average WER between two sequences of morphemes from true transcripts and automatically recognized transcripts was then calculated and shown in Table \ref{table:ling}. The result shows that \textbf{VCTTS-V} outperforms \textbf{VC}, and WER of \textbf{VCTTS-T} is worse than \textbf{TTS}.

After training, eight native Korean speakers participated in subjective evaluation. 20 sentences were generated by \textbf{VC} and \textbf{VCTTS-V} models. Emotion of $\textbf{x}_c$ was set to neutral and target emotion was set to happiness while sentence of $\textbf{x}_s$ is fixed. It is blindly tested that subjects never knew which model generated which speeches. Subjects were asked to rate its clarity from 1 to 5. Also, preference ABX test between two models were conducted. Given two speeches without information that which model generated which samples, subjects were asked to choose clearer speech. Subjects could choose nothing if two samples are perceived similar. The results are shown in Table \ref{table:mosabx}. It shows that \textbf{VCTTV-V} has higher MOS and ABX preference score, which means multitask learning of VC and TTS is helpful to keep linguistic information.

\subsection{Emotional voice conversion}

\begin{figure}[t]
    \vspace{-0.6cm}
    \centerline{
    \includegraphics[width=8.5cm]{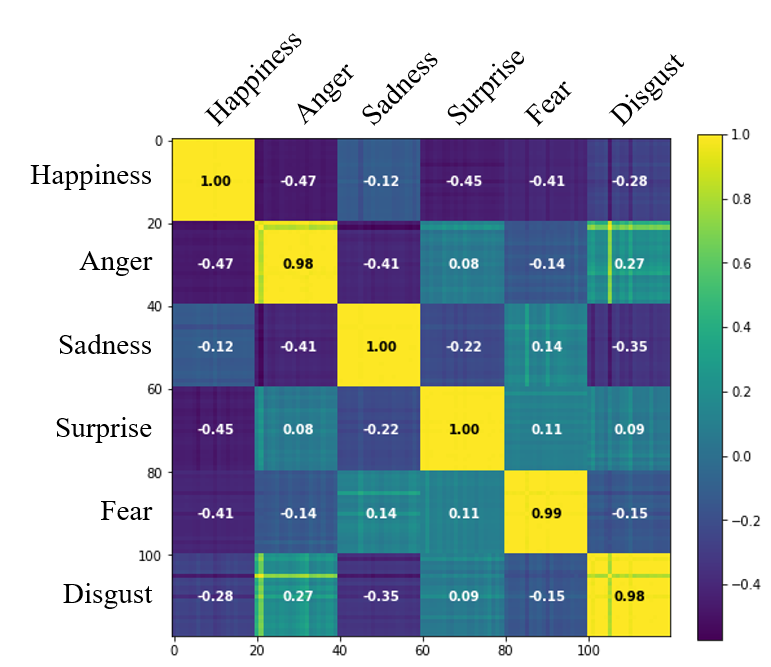}
    }
    \caption{Confusion matrix of the style vector. Mean value of the similarity between emotion pairs are written in each pairs.}
    \label{fig:confusion}
\end{figure}

\begin{figure}[t]
    \centerline{
    \includegraphics[width=8.5cm]{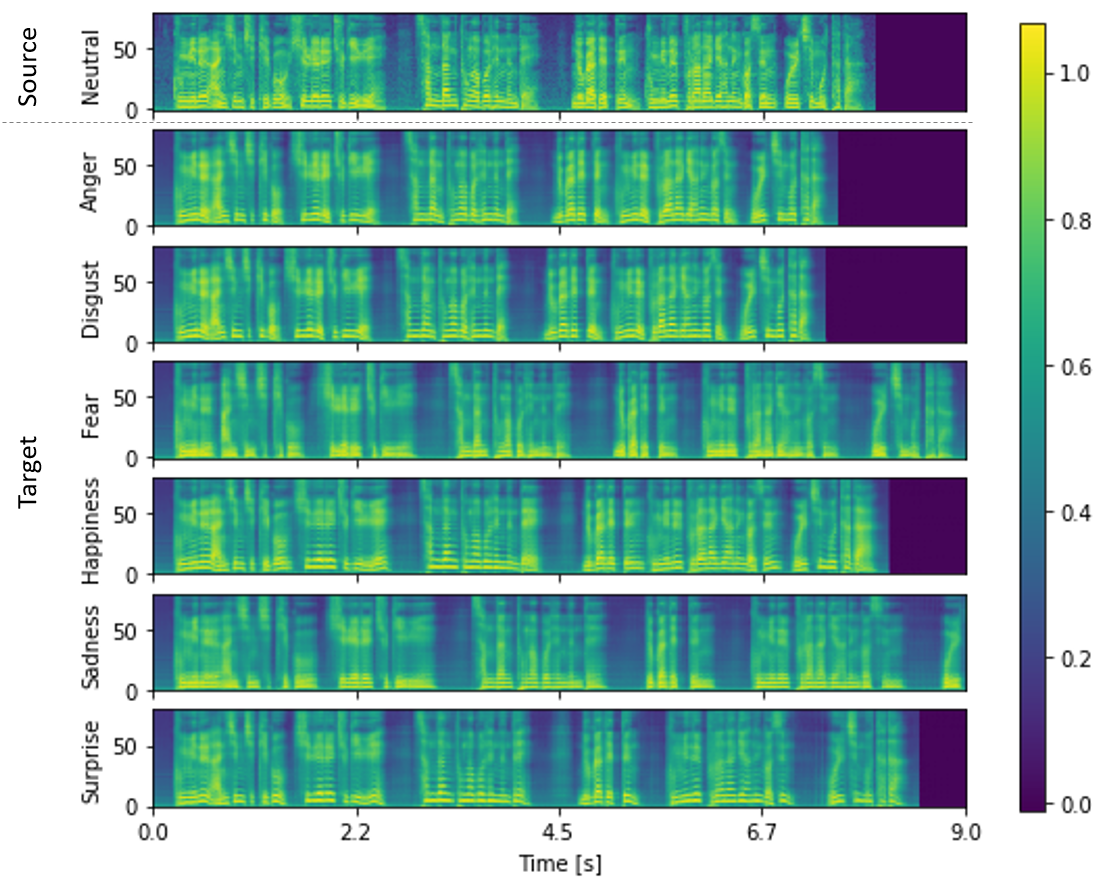}
    }
    \caption{Mel spectrogram of source speech (top row) and converted speeches (other rows).}
    \label{fig:mels}
\end{figure}

To investigate emotional voice conversion, \textbf{VCTTS} model mentioned above is used for inference. After training, we randomly chose 20 samples per each emotion, and those samples are fed to the model. Then $\textbf{h}_s$ can be obtained per sample, and cosine similarity between each sample is measured and illustrated in Fig. \ref{fig:confusion}. The mean value of cosine similarity between all emotion pairs are also shown.

In the figure, it is shown that samples in the same emotion have very high cosine similarity while off-diagonal emotion pairs have low similarity. It implies that the style encoder is able to extract emotion style regardless of linguistic contents. Except for diagonal emotion pairs, emotion pair (Disgust-Anger) also shows relatively high similarity, it means that the embedding of these two emotions is closer than the other emotions. Same phonomenon is observed in (Sadness-Fear) pair.

On the other hand, emotional voice conversion should reflect emotion while keeping linguistic contents. In Fig. \ref{fig:mels}, voice conversion examples are displayed. Given a neutral speech, it is transformed into six different emotions with given $\x_s$. Contents of $\x_s$ is fixed for this experiment. The top row in the Fig. \ref{fig:mels} is log Mel spectrogram of input speech and from the second row to the seventh row are log Mel spectrogram of the converted speech. It can be found that the overall shape of the spectrogram is similar to that of input, while some changes such as temporal shift, frequency shift, duration of pause were made. Within a single model, it can generate multiple domains of speech.

\section{Conclusion}
\label{sec:conclusion}

In this paper, we presented emotional VC using multitask learning with TTS. Although there have been abundant researches on VC, the performance of VC lacks in terms of preserving linguistic information, emotional, and many-to-many VC. Unlike previous methods, the linguistic contents of VC are preserved by multitask learning with TTS. A single model is trained to optimize on both VC and TTS, the embedding space is trained to capture linguistic information by TTS. The result shows that using multitask learning much reduces WER, and subjective evaluation also supports this. For emotional VC, we collected a Korean parallel database for seven distinct emotions, and the model is trained to generate speech depends on the style reference input. Also, style encoder is devised to extract style information while removing linguistic information. Without explicit input of emotion label, style encoder successfully disentangles emotions. 

This research can be extended to many other directions. First, we only show the possibility of helping TTS to VC, but TTS can be also improved by VC since some language has highly nonlinear relationship between text and its pronunciation. Second, the contents encoder is trained to extract only linguistic information, it could be extended to improve speech recognition. Third, more explicit loss such as domain adversarial loss can be added to minimize the difference between embedding of VC and TTS in linguistic space. 

\vfill
\newpage

\bibliographystyle{IEEEbib}
\bibliography{refs}

\end{document}